\documentclass[twocolumn,preprintnumbers,elsart]{revtex4}
\usepackage{amssymb}
\usepackage{amsmath}
\usepackage{graphicx}
\usepackage{mathrsfs}
\usepackage{dcolumn}
\usepackage{color}
\usepackage[center]{subfigure}
\usepackage{bm}

\begin{document}

\title{Two-dimensional vortex quantum droplets}
\author{Yongyao Li$^{1}$, Zhaopin Chen$^{2}$, Zhihuan Luo$^{3}$, Chunqing
Huang$^{1}$, Haishu Tan$^{1}$, Wei Pang$^{4}$}
\email{kingprotoss@gmail.com}
\author{Boris A. Malomed$^{2,1}$}
\affiliation{$^{1}$School of Physics and Optoelectronic Engineering, Foshan University,
Foshan 528000, China \\
$^{2}$ Department of Physical Electronics, School of Electrical Engineering,
Faculty of Engineering, Tel Aviv University, Tel Aviv 69978, Israel.\\
$^{3}$College of Electronic Engineering, South China Agricultural
University, Guangzhou 510642, China\\
$^{4}$Department of Experiment Teaching, Guangdong University of Technology,
Guangzhou 510006, China}

\begin{abstract}
It was recently found that the Lee-Huang-Yang (LHY) correction to the
mean-field Hamiltonian of binary atomic boson condensates suppresses the
collapse and creates stable localized modes (two-component \textquotedblleft
quantum droplets", QDs) in two and three dimensions (2D and 3D). In
particular, the LHY effect modifies the effective Gross-Pitaevskii equation
(GPE) in 2D by adding a logarithmic factor to the usual cubic term. In the
framework of the accordingly modified two-component GPE system, we construct
2D\ self-trapped modes in the form of QDs with vorticity $S$ embedded into
each component. Due to the effect of the logarithmic factor, the QDs feature
a flat-top shape, which expands with the increase of $S$ and norm $N$. An
essential finding, produced by a systematic numerical investigation and
analytical estimates, is that the vortical QDs are \emph{stable} (which is a
critical issue for vortex solitons in nonlinear models) up to $S=5$, for $N$
exceeding a certain threshold value, which is predicted to scale as $N_{%
\mathrm{th}}\sim S^{4}$ for large $S$ (for three-dimensional QDs, the
scaling is $N_{\mathrm{th}}\sim S^{6}$). The prediction is corroborated by
numerical findings. Pivots of QDs with $S\geq 2$ are subject to structural
instability, as specially selected perturbations can split the single pivot
in a set of $S$ or $S+2$ pivots \ corresponding to unitary vortices;
however, the structural instability remains virtually invisible, as it
occurs in a broad central \textquotedblleft hole" of the vortex soliton,
where values of fields are very small, and it does not cause any dynamical
instability. In the condensate of $^{39}$K atoms, in which QDs with $S=0$
and a quasi-2D shape were created recently, the vortical droplets may have
radial size $\lesssim 30$ $\mathrm{\mu }$m, with the number of atoms in the
range of $10^{4}-10^{5}$. The role of three-body losses is considered too,
demonstrating that they do not prevent the creation of the vortex droplets,
but may produce a noteworthy effect, leading to sudden splitting of
\textquotedblleft light" droplets. In addition, \textit{hidden-vorticity}
states in QDs, with topological charges $S_{+}=-S_{-}=1$ in their
components, which are prone to strong instability in other settings, have
their stability region too. Unstable HV states tend to spontaneously merge
into zero-vorticity solitons. Collisions of QDs, which may lead to their
merger, and dynamics of elliptically deformed QDs (which form rotating
elongated patterns or ones with oscillations of the eccentricity) are
briefly considered too.
\end{abstract}

\maketitle

\section{Introduction}

The mean-field approximation \cite{GP1,GP2} offers an extremely accurate
description of one-, two, and three-dimensional (1D, 2D, 3D) matter-wave
patterns in atomic Bose-Einstein condensates (BECs), provided that the
multidimensional states are not made unstable by the occurrence of the
collapse \cite{review}. However, the usual cubic nonlinearity, which
represents attractive interactions between atoms, gives rise to the critical
and supercritical collapse in the 2D and 3D geometries, respectively \cite%
{Berge',Fibich}, which makes search for physically relevant mechanisms
stabilizing 2D and 3D solitons and solitary vortices a challenging problem
\cite{review,review2}. If the physical setting admits the inclusion of
repulsive quintic nonlinearity in addition to the cubic self-focusing, the
balance of these terms stabilizes 2D and 3D fundamental solitons, with
vorticity $S=0$, as well as a part of 2D and 3D vortex-soliton families,
with $S\geq 1$ \cite{Spain,Pego,Kiev} and $S=1$ \cite{3D-vortex},
respectively. The stabilization of fundamental 2D solitons by the
cubic-quintic nonlinearity has been experimentally demonstrated in optics
\cite{Cid}, while quasi-stable solitary vortices were demonstrated only as
transient modes, using the instability-suppressing effect of the cubic loss
(two-photon absorption) \cite{Cid2}. However, the relevance of the
cubic-quintic nonlinearity in the context of BEC is doubtful. A recently
found possibility to create 2D vortex solitons, which are \emph{stable}, at
least, up to $S=5$, was predicted in the framework of the binary BEC\ with
its two components coupled by a microwave field \cite{Jqing2016}.
Stabilization of higher-order vortices with $S>1$ is an especially
interesting issue in this and other contexts.

It was also recently found that the spin-orbit coupling in binary BECs with
the cubic attraction makes otherwise unstable 2D solitons an absolutely
stable ground state \cite{BenLi,Luca,Yongyao2017}, and creates metastable
solitons in 3D \cite{HanPu,SKA}. However, all modes with overall vorticity $%
S\geq 1$ added to them are completely unstable in the spin-orbit-coupled
system \cite{BenLi}. Thus far, no truly stable bright vortex solitons have
been created experimentally in any uniform physical medium or in free space
\cite{review,review2} (it may be easier to create 3D modes with embedded
vorticity in deeply structured media, such as quasi-discrete ``optical
bullets" in a 2D waveguiding array \cite{Jena}).

A new approach to the creation of stable self-trapped 2D and 3D states in
the BEC was recently elaborated in Refs. \cite{Petrov2015,Petrov2016} in a
binary BEC, with self-repulsion in each component and dominating
inter-component attraction. In this setting, the stability against the
collapse is provided by the Lee-Huang-Yang (LHY) correction to the
mean-field\ dynamics, originating from quantum fluctuations around the
mean-field states \cite{LHY1957}. In terms of the respective 3D
Gross-Pitaevskii equations (GPEs), the LHY correction is represented by
quartic self-repulsive terms, which arrest the collapse induced by the
cross-attraction between the components \cite{Petrov2015} (or by the
three-body attraction in 1D \cite{Sekino2018}). The analysis predicts stable
\textit{quantum droplets} (QDs) of an ultradilute superfluid, while
traditional solitons in BEC were created as quantum-gas clouds \cite{Randy}-%
\cite{Aspect}. A variant of this system, which includes the linear Rabi
coupling between the components, was considered too \cite{Luca2}. The
interplay of the LHY and spin-orbit-coupling effects was addressed in Refs.
\cite{Zhengwei2013}, \cite{Yongyao2017}, and \cite{new} (the latter work
applied these ingredients to a Bose-Fermi mixture).

Beyond the framework of the corrected mean-field theory, the zero-vorticity
QDs and their stability were recently explored, by means of the quantum
Monte-Carlo method in Ref. \cite{Boronat}, and the stability of uniform
media filling broad \textquotedblleft droplets" was studied, by means of
nonperturbative treatment of correlations, in Ref. \cite{Zillich}. Further,
beyond-the-mean-field theory valid at finite temperature was recently
developed too \cite{Chiquillo}.

QDs have been created in the binary BEC composed of $10^{4}-10^{5}$ atoms of
$^{39}$K, kept in two different states, with appropriate signs of the inter-
and intra-component interactions \cite{Cabrera2017}-\cite{Inguscio} (the
corresponding scattering lengths in $^{39}$K may be adjusted by the Feshbach
resonance \cite{FR1,FR2}). In particular, in works \cite{Cabrera2017} and
\cite{Cabrera2} the QDs were produced \ with a quasi-2D (\textquotedblleft
pancake") shape, imposed by the strong confining potential applied
perpendicular to the pancake's plane. On the other hand, the QDs reported in
Ref. \cite{Inguscio} were nearly isotropic (spherically symmetric), as they
were not essentially compressed by the confining potential.

Another possibility for the creation of QDs was realized in dipolar BEC,
where the balance of attractive dipole-dipole interactions and LHY repulsion
makes it possible to create single-component QD states, as demonstrated,
using $^{164}$Dy and $^{166}$Er condensates, in Refs. \cite{Barbut2016} and
\cite{Schmitt2016}, respectively. The dynamics of QDs in dipolar BECs\ was
analyzed in detail in recent works \cite{PRX}-\cite{Sadhan}.

The availability of QDs in the current experiments suggests to analyze a
possibility of the creation of such states with embedded vorticity. In
single-component dipolar condensates, the vortex states were recently found
to be unstable, because deformation of the vortex embedded in the QD may
lower its energy \cite{Macri}. On the other hand, it was demonstrated, also
very recently, that two-component QDs supported by the contact interactions
in 3D may readily carry stable embedded vorticities with winding number $S=1$%
, and the QDs with very large norms may also support stable embedded
vorticity $S=2$ \cite{Leticia}.

The objective of the present work is to construct families of stable QDs
with embedded vorticity in the effectively 2D setting. The reduction of the
dimension from 3 to 2 changes the form of the nonlinearity in the underlying
Gross-Pitaevskii equations (GPEs) including the LHY correction \cite%
{Petrov2016}, and thus gives rise to a specific model in the framework of
which the analysis produces vortex solitons. A remarkable fact is that
conspicuous stability regions are found for the solitons with vorticities up
to $S=5$.

The paper is structured as follows. The model is introduced in Sec. II.
Basic results for QDs with explicit and \textquotedblleft hidden"
vorticities are presented in Sec. III [\textquotedblleft hidden" means that
two components of the QD have equal norms and opposite vorticities, see Eq. (%
\ref{+-1}) below]. In addition to systematically collected numerical
findings, essential analytical results are reported too. They include
estimates for the size of the inner \textquotedblleft hole" of the QDs,
induced by the vorticity, and threshold (minimum) value of the norm
necessary for the stability of the vortex QDs, in the form of $N_{\mathrm{th}%
}\sim S^{4}$ [for three-dimensional vortex QDs, it is replaced by $N_{%
\mathrm{th}}(S)\sim S^{6}$]. Another feature predicted by the analytical
approach and confirmed numerically is the effect of the structural
instability of the multiple vortices, with $S>1$: while the vortex QD as a
whole remains stable, small perturbations, with their own intrinsic
vorticities $s=+1$ and $s=-1$, split the pivot of the multiple vortex in a
set of unitary ones, with the number of secondary pivots being,
respectively, $S$ or $S+2$. The split pivots stay close to each other,
without initiating the growth of a dynamical instability. Estimates for
experimentally relevant parameters of the predicted stable modes, along with
analysis an effect of cubic loss, which may also be an experimentally
relevant factor, are given in Sec. V. Collisions between QDs, as well as
dynamics of elliptically deformed ones, are briefly addressed in Sec. VI.
The paper is concluded by Sec. VII.

\section{The model}

The binary BEC in 3D is governed by the GPE system with the cubic terms,
supplemented by the above-mentioned LHY-induced quartic self-repulsion \cite%
{Petrov2016}. For the quasi-2D BEC, strongly confined in the transverse
direction, the GPE-LHY system reduces to a 2D form \cite{Petrov2016}, which
essentially simplifies for modes with lateral size $l\gg \sqrt{a_{\pm
}a_{\perp }}$, where $a_{\pm }$ and $a_{\perp }$ are, respectively, the
self-repulsion scattering lengths of each component, and the
transverse-confinement length \cite{Yongyao2017}. This condition definitely
holds for values relevant to the current experiments, $l\sim 10$ $\mathrm{%
\mu }$m, $a_{\pm }\sim 3$ nm, $a_{\perp }\lesssim 1$\ $\mathrm{\mu }$m \cite%
{Cabrera2017,Inguscio,Cabrera2}, leading to the reduced 2D system for scaled
wave functions $\psi _{\pm }$ of the two components \cite%
{Petrov2016,Yongyao2017}, with coordinates and time additionally rescaled by
$\left( x,y\right) \rightarrow \left( g/2\sqrt{\pi }\right) \left(
x,y\right) ,t\rightarrow \left( g^{2}/4\pi \right) t$:
\begin{gather}
i\frac{\partial \psi _{\pm }}{\partial t}=-\frac{{1}}{2}\nabla ^{2}\psi
_{\pm }+\frac{4\pi }{g}(|\psi _{\pm }|^{2}-|\psi _{\mp }|^{2})\psi _{\pm }
\notag \\
+\left( |\psi _{+}|^{2}+|\psi _{-}|^{2}\right) \psi _{\pm }\ln (|\psi
_{+}|^{2}+|\psi _{-}|^{2})-iL_{3}|\psi _{\pm }|^{4}\psi _{\pm },
\label{Fulleq}
\end{gather}%
where $g>0$ is the coupling constant, and the symmetry between the
components is assumed, $a_{+}=a_{-}$. In the experimental situation, the
latter condition holds only approximately (in particular, in the setting
reported in Ref. \cite{Cabrera2017}, the relative difference between $a_{+}$
and $a_{-}$ is $\simeq 20\%$ at magnetic fields close to $55.5$ G). Here, we
focus on the symmetric system as the simplest one which may be close enough
to the experiment (cf. Refs. \cite{Petrov2015} and \cite{Petrov2016}), while
the detailed analysis of the asymmetric system should be a subject for a
separate work. Further, $L_{3}$ in Eq. (\ref{Fulleq}) is the scaled rate of
three-body losses, which should be taken into account in the realistic
situation \cite{Cabrera2017,Inguscio,Cabrera2}.

For symmetric states with
\begin{equation}
\psi _{+}=\psi _{-}\equiv \psi /\sqrt{2},  \label{+-}
\end{equation}%
system (\ref{Fulleq}) admits reduction to a single equation,
\begin{equation}
i\frac{\partial \psi }{\partial t}=-\frac{{1}}{2}\nabla ^{2}\psi +|\psi
|^{2}\psi \ln (|\psi |^{2})-i\frac{L_{3}}{4}|\psi |^{4}\psi .  \label{GPE}
\end{equation}%
The symmetric states are characterized by the total norm, $N=\int \int |\psi
(\mathbf{r})|^{2}dxdy$ (it is subject to slow decay under the action of the
loss, if the latter is present). In the case of $L_{3}=0$, the Hamiltonian
(energy) corresponding to Eq. (\ref{GPE}) is%
\begin{equation}
E=\frac{{1}}{2}\int \int \left[ \left\vert \nabla \psi \right\vert
^{2}+|\psi |^{4}\ln \left( \frac{|\psi |^{2}}{\sqrt{e}}\right) \right] dxdy
\label{E}
\end{equation}%
($e$ is the base of natural logarithms). Equation (\ref{GPE}) with $L_{4}=0$
also conserves the linear and angular momenta: $\mathbf{P}=i\int \int \psi
^{\ast }\nabla \psi dxdy$, and%
\begin{equation}
M=i\int \int \psi ^{\ast }\left( y\frac{\partial \psi }{\partial x}-x\frac{%
\partial \psi }{\partial y}\right) dxdy\equiv i\int \int \psi ^{\ast }\frac{%
\partial \psi }{\partial \theta }dxdy,  \label{M}
\end{equation}%
where $\ast $ stands for the complex conjugation, and $\left( r,\theta
\right) $ is the set of the polar coordinates in the $\left( x,y\right) $
plane.

Our first objective is to construct 2D self-trapped modes with embedded
vorticity, and explore their stability, in the lossless model with $L_{3}=0$
(previously, only zero-vorticity 2D solutions, whose stability is obvious,
were addressed in the context of the QD models \cite{Petrov2016}); then, the
effect of the three-body loss will be considered. This objective is relevant
as the droplets created in recent works \cite{Cabrera2017,Cabrera2} feature
a strongly oblate form, with the ratio of the radial and transverse sizes $%
\gtrsim 10$.

QD solutions to Eq. (\ref{GPE}), with vorticities $S=0,1,2,...$ , are looked
for, in the polar coordinates, as
\begin{equation}
\psi (\mathbf{r},t)=\phi (r)\exp \left( -i\mu t+iS\theta \right) ,
\label{psiphi}
\end{equation}%
where $\mu <0$ is a chemical potential, and real amplitude function $\phi
(r) $ obeys a radial equation (with $L_{3}=0$),
\begin{equation}
\mu \phi =-\frac{1}{2}\left( \frac{d^{2}\phi }{dr^{2}}+\frac{1}{r}\frac{%
d\phi }{dr}-\frac{S^{2}}{r^{2}}\phi \right) +\phi ^{3}\ln \left( \phi
^{2}\right) .  \label{phi}
\end{equation}%
Note that, as it follows from Eq. (\ref{M}), the angular momentum of QD (\ref%
{psiphi}) (its ``spin momentum") is proportional to its integral norm,%
\begin{equation}
M_{\mathrm{spin}}=SN.  \label{spin}
\end{equation}%
The same equation (\ref{phi}) describes the radial structure of
two-component QDs with \textit{hidden vorticity} (HV), in the form of $\psi
_{\pm }(\mathbf{r},t)=\phi (r)\exp \left( -i\mu t\pm iS\theta \right) $ \cite%
{hidden}, although their stability is completely different, as shown below,
and their total angular momentum is zero (hence the name of ``hidden
vorticity").

Localized solutions of Eq. (\ref{phi}) with fixed $N$ and $S$ were obtained
by means of the imaginary-time-integration method (which may work for vortex
configurations as well as for ground states \cite{IT1,IT2}), applied to Eq. (%
\ref{GPE}) with $L_{3}=0$ and input $\phi _{0}(\mathbf{r})=Cr^{S}\exp
(-\alpha r^{2}+iS\theta ),$ $\alpha >0$. The stability of the stationary
modes was analyzed by means of the linearized Bogoliubov - de Gennes (BdG)
equations \cite{GP1,GP2} for perturbed wave functions, taken as
\begin{equation}
\psi _{\pm }=\left[ \phi _{\pm }(r)+w_{\pm }e^{-i\Lambda t+im\theta }+v_{\pm
}^{\ast }e^{i\Lambda ^{\ast }t-im\theta }\right] e^{-i\mu t+iS\theta },
\label{pert}
\end{equation}%
where $w_{\pm }$ and $v_{\pm }$ are perturbation eigenmodes with integer
azimuthal index $m$, and the imaginary part of eigenvalue $\Lambda $
determines the instability growth rate, if any (see, e.g., Ref. \cite{VS}).
Note that perturbations which break the symmetry of the underlying states
[see Eq. (\ref{+-})] are admitted by Eq. (\ref{pert}). The system of BdG
equations was derived in a straightforward manner, by the substitution of
perturbed wave functions (\ref{pert}) in Eq. (\ref{Fulleq}) and subsequent
linearization. Numerical solution of the linearized equations produces a
spectrum of eigenfrequencies $\Lambda $, the stability condition being that
they all must be real. The so predicted (in)stability was then verified by
direct simulations of the perturbed evolution in the framework of Eq. (\ref%
{Fulleq}), again admitting perturbations which may break the symmetry
relation imposed by Eq. (\ref{+-}).

\section{Results}

\subsection{Zero-vorticity solitons and the flat-top state}

\begin{figure*}[t]
{\includegraphics[width=1.4\columnwidth]{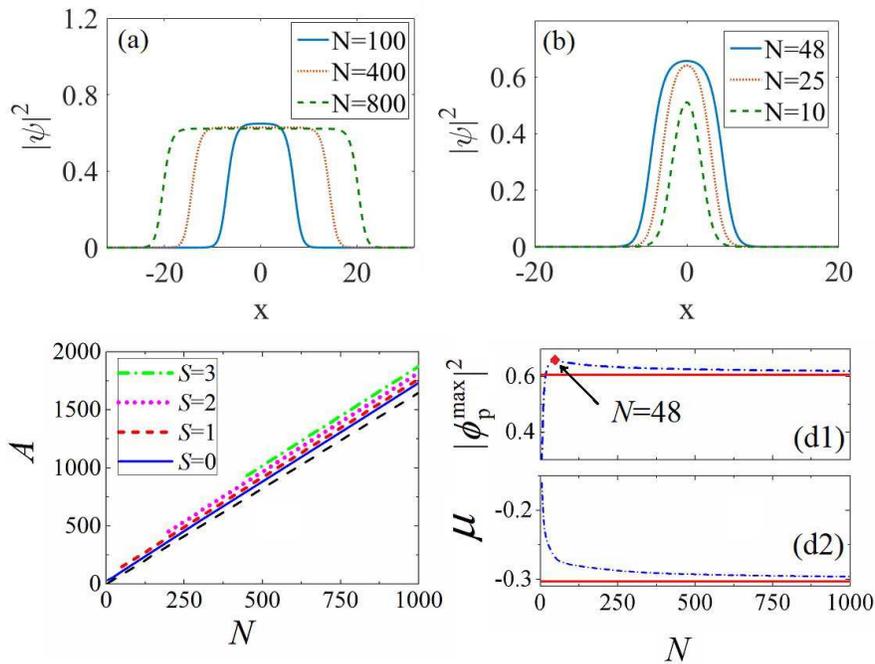}}
\caption{(a) Cross-sections of density patterns $|\protect\phi (\mathbf{r}%
)|^{2}$ for the QDs of the flat-top type with $S=0$ (the system's ground
state) and norms $N=100$ (solid curve), $400$ (dot curve), and $800$ (dash curve). (b) Cross-sections of density
patterns $|\protect\phi (\mathbf{r})|^{2}$ for the QDs of the Gaussian type
with $S=0$ (the system's ground state) and norms $N=10$ (dash curve), $25$ (dot curve), and $48$ (solid curve). (c)
The effective\ area (\protect\ref{Radius}) for QDs with embedded vorticities
$S=0$ (solid line) ,1 (short dash line),2 (dot line),3 (dash-dot line), vs. $N$. The long dashed line represents the TF prediction, $%
A=N/n_{\mathrm{p}}^{\mathrm{(TF)}}$ for $S=0$, see Eq. (\protect\ref{A}).
(d1,d2) The largest (peak) local density, $|\protect\phi _{\mathrm{p}%
}|^{2}\equiv n_{\mathrm{p}}$ [dash-dot curve in panel (d1)], and chemical potential $\protect\mu $ [dash-dot curve in penal (d2)] vs. $N$.
Solid lines in these two panels show the TF limit values, $n_{\mathrm{p}}^{\mathrm{(TF)}}=1/%
\protect\sqrt{e}$ and $\protect\mu _{\mathrm{TF}}=-1/\left( 2\protect\sqrt{e}%
\right) $, see Eqs. (\protect\ref{0.60}) and (\protect\ref{0.30}), respectively. The 
rhombus in panel (d1) designates the border between the QD profiles of the
Gaussian and flat-top types. }

\label{Fundamental}
\end{figure*}

The analysis demonstrates that all the QDs with $S=0$, which represent the
ground state of the system, are stable. Similar to the situation with 1D QDs
\cite{Grisha}, they feature quasi-Gaussian and \textit{flat-top} shapes at
relatively small and larger norms ($N<48$ \ and $N>48$, respectively), as
shown in Fig. \ref{Fundamental}. Their effective area,
\begin{equation}
A\equiv \left[ \int |\phi (\mathbf{r})|^{4}d\mathbf{r}\right] ^{-1}\left(
\int |\phi (\mathbf{r})|^{2}d\mathbf{r}\right) ^{2},  \label{Radius}
\end{equation}%
is displayed, as a function of norm $N$, along with the peak local density, $%
|\psi \left( \mathbf{r}=0\right) |^{2}$, and chemical potential $\mu $, in
Fig. \ref{Fundamental}. The border between the quasi-Gaussian and flat-top
shapes [designated by the the red rhombus at $N=48$ in Fig. \ref{Fundamental}%
(d1)] corresponds to the maximum of the peak density. In this and other
figures,the length and time units correspond to $\sim 1$ $\mathrm{\mu }$m
and $\sim 1$ ms respectively, with $N=100$ corresponding to $\sim 5000$
atoms, in terms of the $^{39}$K condensate \cite{Cabrera2017,Cabrera2}. Note
that the $\mu (N)$ curve satisfies the necessary stability condition in the
form of the Vakhitov-Kolokolov criterion \cite{VK,Berge',Fibich}, $d\mu /dN<0
$.

The flat-top structure of fundamental and vortical solitons (for norms which
are not too small) is demonstrated by models which include competing
self-attraction and repulsion terms, such as the nonlinear Schr\"{o}dinger
equations with cubic-quintic \cite{Pushka,Enns,Spain,Pego}, cubic-quartic
\cite{Leticia} and quadratic-cubic \cite{Grisha} nonlinearities. This
feature is demonstrated by the solitons in one \cite%
{Pushka,Enns,Grisha,Luca3}, two \cite{Spain,Pego} and three \cite{Leticia}
dimensions alike. However, in the present system the reason is different, as
Eq. (\ref{phi}) includes a single nonlinear term, $\phi ^{3}\ln \left( \phi
^{2}\right) $. This term switches its sign from attraction to repulsion due
to the reversal of the sign of $\ln \left( \phi ^{2}\right) $ as its
argument changes from $\phi ^{2}<1$ to $\phi ^{2}>1$. This mechanism arrests
the \textit{critical collapse} \cite{Berge',Fibich} in the present 2D
system, driven by the cubic self-attraction, and, thus, it suggests the
possibility of the existence of stable solitons.

The flat-top states may be analyzed, first, by means of the Thomas-Fermi
(TF) approximation, which neglects the derivatives in Eq. (\ref{phi}) with $%
S=0$. The respective energy density of the flat field, which carries the
peak (largest) density, $n_{\mathrm{p}}\equiv |\phi _{\mathrm{p}}|^{2}$, of
the QD profile [see Fig. \ref{Fundamental}(a)] is $\epsilon (n_{\mathrm{p}%
})=n_{\mathrm{p}}^{2}\ln (n_{\mathrm{p}}/\sqrt{e})$, as per Eq. (\ref{E}).
With area
\begin{equation}
A\approx N/n_{\mathrm{p}}  \label{A}
\end{equation}%
of the flat-top QD, its bulk energy, which dominates the total energy (\ref%
{E}), is
\begin{equation}
E_{\mathrm{bulk}}\approx A\epsilon (n_{\mathrm{p}})=Nn_{\mathrm{p}}\ln (n_{%
\mathrm{p}}/\sqrt{e}).  \label{bulk}
\end{equation}%
Then, $n_{\mathrm{p}}$ is determined by the minimization of the energy for
the given norm: $dE/dn_{\mathrm{p}}=0$, yielding
\begin{equation}
n_{\mathrm{p}}^{\mathrm{(TF)}}=1/\sqrt{e}\approx 0.6065,  \label{0.60}
\end{equation}%
which is very close to the numerically found values, see Fig. \ref%
{Fundamental}(c) [and Fig. \ref{propagation}(b1) below]. The respective
chemical potential is
\begin{equation}
\mu _{\mathrm{TF}}=n_{\mathrm{p}}^{\mathrm{(TF)}}\ln (n_{\mathrm{p}}^{%
\mathrm{(TF)}})=-1/\left( 2\sqrt{e}\right) \approx -0.3033,  \label{0.30}
\end{equation}%
in agreement with numerical data displayed in Figs. \ref{Fundamental}(d2)
and \ref{propagation}(b2).

The same results for the peak density and chemical potential can also be
obtained in a different way. Indeed, for broad flat-top QDs, radial equation
(\ref{phi}) becomes quasi-one-dimensional,
\begin{equation}
\mu \phi =-\frac{1}{2}\frac{d^{2}\phi }{dr^{2}}+\phi ^{3}\ln \left( \phi
^{2}\right) .  \label{radial}
\end{equation}%
The respective formal Hamiltonian, which remains constant in the course of
the spatial evolution of the field along $r$, is%
\begin{equation}
h=\frac{\mu }{2}\phi ^{2}+\frac{1}{4}\left( \frac{d\phi }{dr}\right) ^{2}-%
\frac{1}{4}\phi ^{4}\ln \left( \frac{\phi ^{2}}{\sqrt{e}}\right) .  \label{h}
\end{equation}%
For localized QDs with $\phi (r=\infty )=0$, one should set $h=0$ in Eq. (%
\ref{h}). Finally, in the limit of very broad QDs the derivative terms in
Eqs. (\ref{radial})\ and (\ref{h}) may be dropped, which yields an algebraic
system, $\mu =\phi ^{2}\ln \left( \phi ^{2}\right) ,~\mu =(1/2)\phi ^{2}\ln
\left( \phi ^{2}/\sqrt{e}\right) $. A solution of this system is identical
to the values given by Eqs. (\ref{0.60}) and (\ref{0.30}).

\subsection{Vortex solitons}

\subsubsection{The shape}

\begin{figure*}[t]
{\includegraphics[width=1.5\columnwidth]{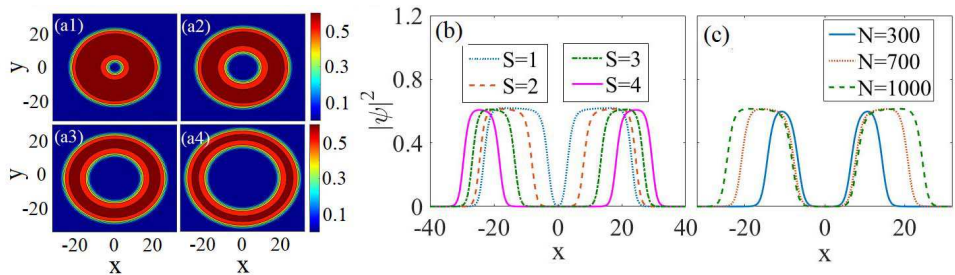}}
\caption{Panels (a1)-(a4) display density patterns of vortex QDs with $S=1$,$%
2$, $3$, $4$ and norm $N=1000$. The first three QDs are stable, while the one
corresponding to $S=4$ is not, see Fig. \protect\ref{propagation}(a2). (b)
Cross-sections of the density patterns from panels (a1)-(a4) (dot, dash, dash-dot, and solid curves represent $S=1$, 2, 3, and 4, respectively). (c) The same
as in (b), for $S=1$ and different values of $N$ (solid, dot and dash curves represent $N=300$, 700, and 1000, respectively). All vortex QDs shown in (c) are stable.}
\label{Vortex1234exp}
\end{figure*}
Typical examples of numerically found vortex QDs with $N=1000$ are displayed
in Figs. \ref{Vortex1234exp}(a,b), which shows that they are flat-top rings,
with inner and outer radii growing with the increase of $S$. The inner
radius can be roughly predicted using the TF approximation (which may be
relevant for vortices \cite{Feder}), applied to Eq. (\ref{phi}). This yields
either $\phi =0$ or%
\begin{equation}
\mu -\phi ^{2}\ln \left( \phi ^{2}\right) =\frac{S^{2}}{2r^{2}}.  \label{TF}
\end{equation}%
Equation (\ref{TF}) shows that $\phi (r)$ keeps decreasing, following the
decrease of $r$ towards $r=0$, up to $\phi _{\min }=1/\sqrt{e}$.
Substituting here $\phi =\phi _{\min }$ and $\mu =\mu _{\mathrm{TF}}$ from
Eq. (\ref{0.30}), we find that $\phi =\phi _{\min }$ is attained at%
\begin{equation}
r_{\min }=\sqrt{e/(2-\sqrt{e})}S\approx 2.8S.  \label{rmin}
\end{equation}%
At $r<r_{\min }$, the TF solution given by Eq. (\ref{TF}) cannot be used,
hence it jumps to $\phi =0$. Thus, $r_{\min }$ can be used as an estimate
for the radius of the inner \textquotedblleft hole" of the vortex soliton.

\begin{figure*}[t]
{\includegraphics[width=1.6\columnwidth]{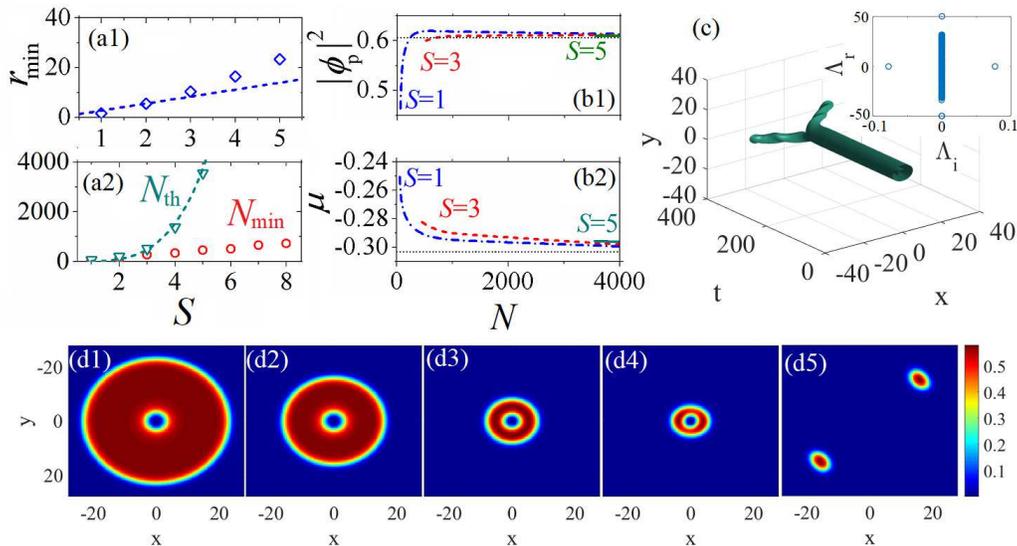}}
\caption{(a1) The numerically found radius of the vortex' inner hole vs. $S$
[each point corresponds to $N_{\mathrm{th}}(S)$ in panel (a2)], the dashed
line showing the prediction given by the TF approximation, according to Eq. (%
\protect\ref{rmin}). (a2) Minimum norms, $N_{\min }$, necessary for the
existence of the vortex QDs (circles), and threshold values of the norm
necessary for their stability, $N_{\mathrm{th}}$ (triangles), vs $S$. The
dashed curve shows the fit to analytically predicted scaling (\protect\ref%
{^4}), $N_{\mathrm{th}}=6S^{4}$. Numerically exact values of $N_{\min }(S)$
and $N_{\mathrm{th}}(S)$ are additionally produced in Table 1. (b1,b2) The
peak density and chemical potential of QDs with different vorticities, vs. $N
$ (dash-dot, dash, and solid curves represent $S=1$, 3, and 5, respectively ). Horizontal dot lines show the TF-predicted limit values $n_{\mathrm{p}}^{\mathrm{(TF)}}$ and $\protect\mu _{\mathrm{TF}}$, see Eqs. (\protect\ref%
{0.60}) and (\protect\ref{0.30}), respectively. (c) The evolution of an
unstable QD with $(N,S)=(30,1)$. The inset shows the respective spectrum of
stability eigenvalues $\Lambda $. The above panels pertain to $L_{3}=0$ (no
three-body loss). (d) The evolution of the density pattern of the QD with $%
S=1$ and initial norm $N=1000$, produced by simulations of Eq. (\protect\ref%
{GPE}) with $L_{3}=0.01$, starting from $t=0$ (d1). Other plots pertain to $%
t=464$ (d2), $1584$ (d3), $3008$ and $3120$ (d4,d5).}
\label{propagation}
\end{figure*}

The area of the vortex QDs, defined as per Eq. (\ref{Radius}), is shown, as
a function of $N$, in Fig. \ref{Fundamental}(c), which implies that it is
well estimated, as above, by relation $A=\sqrt{e}N$, that directly follows
from Eqs. (\ref{A}) and (\ref{0.60}). Further, Figs. \ref{Vortex1234exp} and %
\ref{propagation}(a1) demonstrates that the vortex' internal radius grows
quasi-linearly with $S$ and does not depend on $N$, as predicted by Eq. (\ref%
{rmin}), the numerically found values of the radius being relatively close
to those given by Eq. (\ref{rmin}). Figures \ref{propagation}(b1,b2)
demonstrate that, with the increase of $N$, the peak density of vortex rings
quickly tends to the TF limit value given by Eq. (\ref{0.60}), while the
chemical potential approaches the respective limit, given by Eq. (\ref{0.30}%
), much slower, due to the contribution of the vorticity.

\subsubsection{Dynamical stability: numerical results and analytical
estimates}

To address the stability of the vortices, we first note that, if they are
shaped as relatively narrow annuli (see Fig. \ref{Vortex1234exp}), it is
possible to produce an estimate based on the consideration of the azimuthal
modulational instability (MI)\ in the 1D reduction of Eq. (\ref{Fulleq}) for
the narrow ring; we stress that this consideration admits perturbations
violating the equality between the components, see Eq. (\ref{+-}). The
result is that MI \emph{does not} occur if the density is large enough, $%
n\equiv |\psi |^{2}>1/e\approx 0.6~n_{\mathrm{p}}^{\mathrm{(TF)}}$. Coupling
constant $g$ from Eq. (\ref{Fulleq}) does not appear in this condition,
suggesting that the full stability may be insensitive to $g$. For the HV
states the analysis produces a more restrictive condition for the absence of
MI, which includes $g$: $\left( n-\pi gS^{2}/R^{2}\right) \ln \left(
en\right) >\left( 2\pi S/R\right) ^{2}$, where $R$ is the ring's radius,
suggesting that HV modes are more prone to instability, depending on the
value of $g$. These qualitative predictions are confirmed by numerical
results reported below.

Data produced by the numerical solution of the linearized equations for
eigenmodes of small perturbations, and fully confirmed by direct simulations
of the perturbed evolution (not shown here in detail), reveal that vortex
QDs in the system without the three-body loss exist above a minimum value of
the norm, $N>N_{\min }$ [which is not surprising, as, in the framework of
the cubic GPE, (unstable) 2D solitons with each value of $S$ exist at a
single value of $N$ \cite{Berge',Fibich}], and they are stable above a
certain threshold \ value, i.e., at $N>N_{\mathrm{th}}>N_{\min }$, see Fig. %
\ref{propagation}(a2), which shows that $N_{\mathrm{th}}$ steeply grows with
$S$. As a result, stable vortices were found up to $S=5$, but not for $S\geq
6$. In this connection, it is relevant to mention that the recent analysis
of the three-dimensional QD model has revealed stability solely for $S=1$
and $2$, with an extremely large \ $N_{\mathrm{th}}(S=2)$ \cite{Leticia}. In
addition to Fig. \ref{propagation}(a2), numerically exact values of $N_{\min
}$ and $N_{\mathrm{th}}$ are collected in Table 1.

\begin{tabular}{llllll}
\hline
$S$ & $1$ & $2$ & $3$ & $4$ & $5$ \\ \hline
$N_{\min }$ & $25$ & $100$ & $260$ & $340$ & $450$ \\
$N_{\mathrm{th}}$ & $60$ & $200$ & $510$ & $1380$ & $3550$ \\ \hline
\end{tabular}

\noindent {\small Table 1: numerically exact values of lower boundaries in
terms of the norm, necessary for the existence (}$N_{\min }${\small ) and
stability (}$N_{\mathrm{th}}${\small ) of vortex QDs with winding number }$S$%
{\small . The same data are displayed graphically in Fig. \ref{propagation}%
(a2).}

\smallskip In the interval of $N_{\min }<N<N_{\mathrm{th}}$, unstable
vortices with winding number $S$ split, typically, into $S+1$ fragments, see
an example for $S=1$ (splitting into a pair of fragments, each being,
approximately. a zero-vorticity soliton) in Fig. \ref{propagation}(c).
Splitting is a typical outcome of the evolution of unstable vortex solitons
in previously studied models \cite{Spain}-\cite{Kiev}, \cite{hidden}.

We stress that, although the full GPE-LHY system (\ref{Fulleq}) contains
coupling constant $g$, the stability boundary, $N_{\mathrm{th}}(S)$, does
not depend on $g$ (in agreement with the above-mentioned prediction produced
by the azimuthal-MI analysis), i.e., $N_{\mathrm{th}}(S)$ is determined by
the analysis performed in the framework of Eq. (\ref{GPE}), from which $g$
was scaled out, while perturbations breaking the symmetry constraint (\ref%
{+-}), which reduces Eq. (\ref{Fulleq}) to Eq. (\ref{GPE}), do not introduce
additional instabilities.

The dependence of $N_{\mathrm{th}}$ on $S$ can be explained by means of an
analytical approximation, based on the energy estimate for the vortex QDs of
a large radius, $R$ (i.e., with a large norm, $N$), cf. a similar estimate
for the 3D vortical QDs developed in Ref. \cite{Leticia}. Indeed, the
splitting of the flat-top vortical QD in two or several fundamental
solitons, which also feature the flat-top shape, in the first approximation
does not alter the bulk energy, which is proportional to the conserved total
norm of the solitons, as per Eq. (\ref{bulk}). Then, the stability against
the splitting is determined by the balance of three other terms: the surface
energy, $E_{\mathrm{surf}}$, the vortical (phase-gradient) energy of the
unsplit vortex, $E_{\mathrm{vort}}$, and the kinetic energy of the
splinters, $E_{\mathrm{kin}}$, which is determined by the conservation of
the angular momentum.

First, taking into regard that radius $r_{\min }$ of the inner
\textquotedblleft hole" in the first approximation does not depend on the
outer radius, $R$, the contribution of the hole's edge to $E_{\mathrm{surf}}$
is negligible in comparison with that of the outer edge, for $R\gg r_{\min }$%
. Thus, one concludes that, in the simplest case of the splitting of the
vortical soliton of radius $R$ in two fundamental ones [as in Fig. \ref%
{propagation}(c)], radii of the splinters are determined by the conservation
of the total norm:
\begin{equation}
R_{\mathrm{splinter}}=R/\sqrt{2}.  \label{spl}
\end{equation}%
Then, the resultant splitting-induced \emph{increase} of the surface energy
is%
\begin{equation}
\Delta E_{\mathrm{surf}}=2\pi \sigma \left( \sqrt{2}-1\right) R,
\label{surf}
\end{equation}%
where $\sigma $ is an efficient surface tension, which is a fixed constant
for flat-top solitons with the fixed density, close to one given by Eq. (\ref%
{0.60}).

The vortical-energy term in total Hamiltonian (\ref{E}) may be readily
estimated as%
\begin{equation}
E_{\mathrm{vort}}=\pi \int_{0}^{\infty }\left\vert \frac{\partial \psi }{%
\partial \theta }\right\vert ^{2}\frac{dr}{r}\approx \pi S^{2}n_{\mathrm{p}%
}\ln \left( \frac{R}{r_{\min }}\right) .  \label{vort}
\end{equation}%
Lastly, assuming that the two splinters emerge being separated by distance
equal to their diameter, i.e., $2R_{\mathrm{splinter}}=\sqrt{2}R$, pursuant
to Eq. (\ref{spl}), and they start to move with velocities $\pm V$ in
opposite directions, $V$ can be estimated by equating the respective orbital
momentum, $M_{\mathrm{orbit}}=\sqrt{2}RVN$ (the soliton's dynamical mass is
equal to its norm) to the original ``spin" momentum (\ref{spin}): $%
V=S/\left( \sqrt{2}R\right) $. Thus, the kinetic energy of the emerging
soliton pair is%
\begin{equation}
E_{\mathrm{kin}}=\frac{1}{2}NV^{2}\equiv \frac{\pi }{4}n_{\mathrm{p}}S^{2},
\label{kin}
\end{equation}%
where the norm was substituted by
\begin{equation}
N=\pi R^{2}n_{\mathrm{p}},  \label{Nn}
\end{equation}%
as per Eq. (\ref{A}). Analyzing the energy balance for the possible
splitting of the vortex solitons with large $R$, in the lowest approximation
one may neglect $R$-independent term (\ref{kin}) in comparison with those
given by Eqs. (\ref{surf}) and (\ref{vort}), which are growing functions of $%
R$.

Thus, the stability against the splitting may be predicted for (large)
values of $R$ at which, for given (sufficiently large) $S$, the
splitting-induced increase of the surface energy, estimated by Eq. (\ref%
{surf}), \emph{exceeds} the energy drop due to the disappearance of the
vortical energy, which is given by Eq. (\ref{vort}): $\Delta E_{\mathrm{surf}%
}(R)\geq E_{\mathrm{vort}}(R)$. Substituting here expressions (\ref{surf})
and (\ref{vort}), and taking into regard that $\ln \left( R/r_{\min }\right)
$ is a slowly varying function of its argument, the stability boundary
(threshold value of the norm) corresponds to the following scaling relation:%
\begin{equation}
N\geq N_{\mathrm{th}}\simeq \mathrm{const}\cdot S^{4},  \label{^4}
\end{equation}%
where Eq. (\ref{Nn}) was used to eliminate $R$ in favor of $N$. As shown in
Fig. (\ref{propagation}(a2), Eq. (\ref{^4}), with a properly adjusted value
of $\mathrm{const}$, provides quite an accurate fit to the numerically found
dependence $N_{\mathrm{th}}(S)$. It is worthy to mention that this
consideration is quite general, and the asymptotic scaling given by Eq. (\ref%
{^4}) should apply to other 2D models which support stable vortex solitons
with the flat-top shape. For three-dimensional QDs with embedded vorticity,
a similar approximation, which extends that developed in Ref. \cite{Leticia}
for $S=1$, yields $N_{\mathrm{th}}^{\mathrm{(3D)}}\simeq \mathrm{const}\cdot
S^{6}$. This very steep dependence explains why the stability for
three-dimensional vortex QDs was found only for $S=1$, and also for $S=2$
with extremely large values of $N$, but not for $S>2$.

\begin{figure}[t]
{\includegraphics[width=1.0\columnwidth]{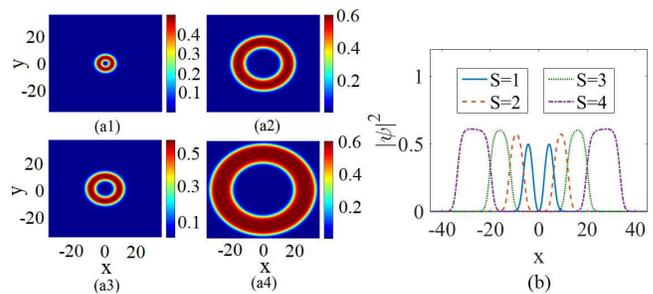}}
\caption{Panels (a1)-(a4) display density patterns of vortex QDs with $S=1$,$%
2$,$3$ and $4$ and norms $N=60$, $200$, $510$ and $1380$, respectively,
which are selected from the stability boundary of the QDs, see Fig. \protect
\ref{propagation}(a2). (b) Cross-sections of the density patterns from
panels (a1)-(a4).}
\label{Vortex1234stable}
\end{figure}

Profiles of vortex QDs with $S=1,2,3,4$, selected precisely at the stability
boundary, are displayed in Fig. \ref{Vortex1234stable}. It is observed that
their shape indeed develops towards the flat-top pattern with the increase
of $S$, cf. typical patterns of the vortex states found in the depth of the
stability area, which are displayed in Fig. \ref{Vortex1234exp} [except for
the state with $S=4$ in Figs. \ref{Vortex1234exp}(a,b), which is still
unstable, as its norm falls slightly below $N_{\mathrm{th}}(S=4)$].

\subsubsection{Structural instability of higher-order vortex solitons}

While, as shown above, solutions for QDs with multiple vorticity, up to $S=5$%
, can be easily found as stable ones, that are actually very robust against
relatively strong perturbations, it is relevant to mention that all the
higher-order vortices, with $S>1$, demonstrate \textit{structural instability%
}, which implies that a specially selected small perturbation, without
causing any dynamical instability, may split the pivot of the $S$-multiple
vortex into sets of $S$ or $S+2$ pivots corresponding to unitary vortices,
although the splitting remains almost invisible, as it occurs in the broad
central \textquotedblleft hole" induced by the multiple vorticity, where
values of the wave fields remain extremely small. In fact, this structural
instability is not specific to the current model, but is quite a generic
effect, therefore we aim to briefly present it here in an explicit form.

In a small vicinity of the pivot of \ a multiple vortex with $S\geq 2$,
placed at the origin ($x=y=0$), a small disturbance with relative amplitude $%
-\varepsilon $ (generally, a complex one), which carries vorticity $s=1$,
may be added, producing a perturbed configuration,%
\begin{eqnarray}
&&\psi _{\mathrm{pert}}\left( x,y\right) \approx \left( x+iy\right)
^{S}-\varepsilon \left( x+iy\right)  \\
&\equiv &\left( x+iy\right) \left[ \left( x+iy\right) ^{S-1}-\varepsilon %
\right] .  \notag  \label{pert1}
\end{eqnarray}%
Pivots of individual (unitary) vortices, into which the small disturbance
splits the multiple vortex, are zeros of $\left\vert \psi _{\mathrm{pert}%
}\left( x,y\right) \right\vert $. \ As seen from Eq. (\ref{pert}), these are
points
\begin{equation}
x_{\mathrm{piv}}^{(1)}=y_{\mathrm{piv}}^{(1)}=0,  \label{1}
\end{equation}%
and $\left( S-1\right) $ additional points, determined as $\left( S-1\right)
$ branches of the root of degree $1/\left( S-1\right) $:
\begin{equation}
x_{\mathrm{piv}}^{(1+j)}+iy_{\mathrm{piv}}^{(1+j)}=\varepsilon ^{1/(S-1)},
\label{j}
\end{equation}%
where $j$ takes values $1,...,S-1$.

Another possibility is to consider a small disturbance with vorticity $s=-1$
(rather than $+1$, as considered above):%
\begin{equation}
\psi _{\mathrm{pert}}\left( x,y\right) =\left( x+iy\right) ^{S}-\varepsilon
\left( x-iy\right) .  \label{pert2}
\end{equation}%
In this case, the pivots are located at points defined by equation $\left(
x+iy\right) ^{S}=\varepsilon \left( x-iy\right) $, i.e., $\left( x+iy\right)
^{S+1}=\varepsilon r^{2}$. After simple manipulations, the latter equation
yields a set of $S+2$ pivots:%
\begin{equation}
x_{\mathrm{piv}}^{(k)}+iy_{\mathrm{piv}}^{(k)}=\varepsilon
^{1/(S+1)}\left\vert \varepsilon \right\vert ^{2/\left( S^{2}-1\right) },
\label{k}
\end{equation}%
where $k$ takes values $1,...,S+1$, plus the central pivot defined by Eq. (%
\ref{1}).

\begin{figure}[t]
{\includegraphics[width=0.8\columnwidth]{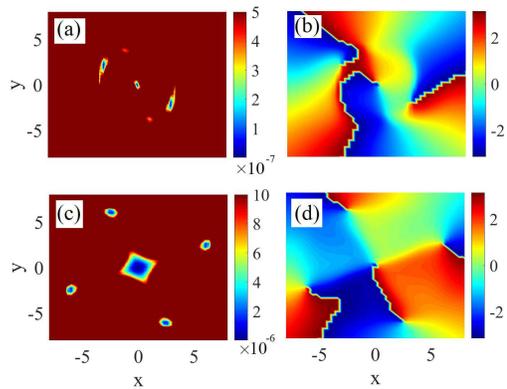}}
\caption{(a) A zoom (in domain $\left\vert x,y\right\vert \leq 8$) of the
density pattern, $\left\vert \protect\psi \left( x,y\right) \right\vert ^{2}$%
, for a QD with $(S,N)=(3,1000)$, which was initially perturbed as per Eq. (%
\protect\ref{pert1}), with $s=+1$ and $\protect\varepsilon =0.0013+0.0023i$.
The pattern is produced by the simulation of Eq. (\protect\ref{GPE}) up to $%
t=5000$. (b) The corresponding phase pattern clearly identifies three
unitary vortices, whose pivots are located at zeros of the local amplitude.
(c) and (d): The same as in (a) and (b), but with the initial perturbation
carrying vorticity $s=-1$, as per Eq. (\protect\ref{pert2}). In this case,
the amplitude and phase patterns demonstrate splitting into a set of five
pivots: one with winding number $-1$ located in the middle, surrounded by
four satellites with winding numbers $+1$. Note the extremely small scale of
the local amplitude, $\left\vert \protect\psi \left( x,y\right) \right\vert
\sim 10^{-7}$ in (a) and (c). On the normal scale, such as one in Figs.
\protect\ref{Vortex1234exp} and \protect\ref{Vortex1234stable}, these
splitting patterns remain invisible.}
\label{pivots}
\end{figure}

These simple arguments were verified by direct simulations, as shown in Fig. %
\ref{pivots}. Strong magnification of numerical data in the nearly empty
area of the central \textquotedblleft hole" precisely confirms the splitting
of the $S$-multiple pivot into $S$ or $S+2$ sets of unitary-vortex pivots,
under the action of the small initial perturbation with its own vorticity $%
s=+1$ or $s=-1$, respectively. Note that the original vortex with $S=1$ is
not subject to the splitting in either case. It is also relevant to stress
that the splitting remains virtually invisible on the normal scale of $%
\left\vert \psi \left( x,y\right) \right\vert $, hence it does not imply any
conspicuous instability of the solitons carrying the multiple vorticity.

\subsection{Hidden-vorticity (HV) modes}

\begin{figure}[t]
{\includegraphics[width=1.0\columnwidth]{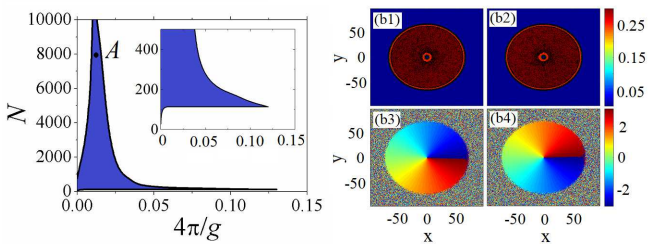}}
\caption{(a) The stability area of HV modes, defined as per Eq. (\protect\ref%
{+-1}), in the plane of $(N,4\protect\pi /g)$, in the model based on Eq. (%
\protect\ref{Fulleq}) with $L_{3}=0$. The indigo (medium gray) area is bounded by the lower and
upper limit values, see Eq. (\protect\ref{NNN}), with nearly constant values
of $N_{\mathrm{th}}^{\mathrm{(low)}}$, as seen in the inset. (b) Density and
phase plots showing opposite vorticities of the two components of a stable
HV mode, with $(N,4\protect\pi /g)=(8000,0.11)$, which corresponds to point
A in (a), as a result of the evolution simulated up to $t=10000$.}
\label{realtime}
\end{figure}

On the contrary to the modes with explicit vorticity, the stability of those
carrying the HV, \textit{viz}.,
\begin{equation}
S_{+}=-S_{-}=1,  \label{+-1}
\end{equation}%
strongly depends on interaction constant $g$ in Eq. (\ref{Fulleq}), making
it necessary to the use the full system, given by Eq. (\ref{Fulleq}), for
the identification of the corresponding stability area, see Fig. \ref%
{realtime}(a). As a result, it is found that the stability is restricted to
sufficiently large values of $g$, \textit{viz}.,
\begin{equation}
4\pi /g<0.13,  \label{1/g}
\end{equation}%
and to values of the norm which are bounded both from above and below:
\begin{equation}
N_{\mathrm{th}}^{\mathrm{(low)}}<N<N_{\mathrm{th}}^{\mathrm{(upp)}},
\label{NNN}
\end{equation}%
with $N_{\mathrm{th}}^{\mathrm{(low)}}\approx 112$ being nearly constant in
almost entire interval (\ref{1/g}), as seen in the inset to Fig. \ref%
{realtime}(a). However, at extremely large values of $g$, namely, $2\pi
/g\lesssim 10^{-3}$, $N_{\mathrm{th}}^{\mathrm{(low)}}$ drops from $112$ to $%
60$. Note that the latter value exactly coincides with the stability
boundary, $N_{\mathrm{th}}$, for the corresponding state with the \emph{%
explicit vorticity}, $S_{+}=S_{-}=1$ [cf. Eq. (\ref{+-1})], as seen in Table
1. Further, Fig. \ref{realtime}(a) identifies an optimum value of the
coupling constant, $g\approx \allowbreak 114$, at which the stability region
of the HV modes extends to extremely large values of the norm. A typical
example of a stable mode, with HV defined as per Eq. (\ref{+-1}), and large $%
N$ is displayed in Fig. \ref{realtime}(b) for $(N,g)=(8000,114)$. A caveat
is that the applicability of the underlying model to so large values of $g$
is not obvious, but the possibility of the existence of stable HV states is
worth noting (in particular, because such stable states were not revealed by
the recent analysis of three-dimensional vortex droplets \cite{Leticia}).
All higher-order HV\ states, with $S_{+}=-S_{-}\geq 2$, were found to be
unstable.

Simulations of the perturbed evolution of unstable HV\ states (not shown
here in detail) demonstrate that vortex cores in their two components split
and start drifting in different directions, being eventually expelled from
the pattern. The final outcome is, in many cases, merger of the unstable HV
state into a zero-vorticity soliton.

\section{Physical estimates}

Translating the scaled units into physical ones (in particular, as per Refs.
\cite{Cabrera2017,Cabrera2}), we conclude that the predicted
vortex modes may have radial size $R$ up to $\lesssim 30$ $\mathrm{\mu }$m
and transverse thickness $\lesssim 1$ $\mathrm{\mu }$m, containing $%
10^{4}-10^{5}$ atoms, with density $\sim 10^{14}$ cm$^{-3}$. The radial size
may be essentially larger than observed in recent experiments for
zero-vorticity oblate droplets \cite{Cabrera2017,Cabrera2}, as they tend to
swell under the action of the embedded vorticity, as seen in Figs. \ref%
{Vortex1234exp}(a,b) and \ref{Vortex1234exp}.

To address the role of the three-body loss, we note that the loss rate in
physical units for $^{39}$K is $\simeq 10^{-27}$ cm$^{6}$/s \cite{Shlyap}.
With the above-mentioned typical densities, $n\sim 10^{14}$ cm$^{3}$, this
implies the decay time $t_{\mathrm{decay}}\simeq 10\left( Kn^{2}\right)
^{-1}\sim 1$ s, which is $t\sim 1000$ in terms of our scaled notation. Then,
the respective estimate for the scaled loss coefficient is $L_{3}\sim 0.01$.
Figure \ref{propagation}(d) displays the simulated evolution, in the
framework of Eq. (\ref{GPE}), of the density pattern for an originally
stable vortex QD with $(S,L_{3})=(1,0.01)$ and initial norm $N_{0}=1000$.
The loss gives rise to shrinkage of the QD, which keeps the flat-top shape,
with the fixed density, $|\psi |^{2}\approx n_{\mathrm{p}}^{\mathrm{(TF)}}$,
as per Eq. (\ref{0.60}), until the slowly decaying norm drops below the
stability threshold, $N_{\mathrm{th}}$ [see Fig.. \ref{propagation}(a2)],
causing quick spitting of the vortex mode. Similarly, sudden splitting into $%
S+1$ fragments, on the experimentally relevant time scale, is caused by the
loss-induced evolution of originally stable droplets with $S>1$, provided
that $N_{0}$ is not too large.

The loss-induced shrinkage of the flat-top droplets can be easily predicted
analytically. As follows from the full equation (\ref{GPE}), the total norm
decays in time as
\begin{equation}
\frac{dN}{dt}=\frac{L_{3}}{2}\int \int |\psi |^{6}dxdy.
\end{equation}%
Further, it follows from here that external radius $R$ of the flat-top QD
with inner density $n_{\mathrm{p}}$ given by Eq. (\ref{0.60}), which is
related to $N$ according to Eq. (\ref{Nn}), shrinks in the course of the
loss-induced evolution as
\begin{equation}
R(t)=R_{0}\exp \left( -\frac{{L_{3}}t}{4e}\right) ,  \label{R(t)}
\end{equation}%
where $R_{0}$ is the initial radius, the corresponding decay law of the norm
being
\begin{equation}
N(t)=\left( \frac{\pi }{\sqrt{e}}\right) R_{0}^{2}\exp \left( -\frac{{L_{3}}t%
}{2e}\right) .  \label{N(t)}
\end{equation}%
Predictions produced by Eqs. (\ref{R(t)}) and (\ref{N(t)}) are well
corroborated by the numerical simulations.

The present analysis is performed in the zero-temperature limit (the
condensation point in the experimentally relevant setting is $T_{\mathrm{BEC}%
}\simeq 150$ nK \cite{K-39}). Finite-$T$ effects for zero-vorticity droplets
were recently studied in Ref. \cite{T}, by means of the
Hartree-Fock-Bogoliubov-Popov \cite{Nick} theory, with a conclusion that the
thermal component forms a halo around the droplets. A similar prediction is
expected for the vortex QDs, with some density of thermal atoms to be found
in the droplet's inner hole too. Detailed studies of the finite-$T$ setting
should be a subject for a separate work.

\section{Interactions between quantum droplets and dynamics of elliptically
deformed ones}

Collisions between effectively one-dimensional QDs were recently analyzed in
Ref. \cite{Grisha}, where both quasi-elastic interactions and strongly
inelastic outcomes, in the form of merger of colliding QDs, were reported.
These results suggest to simulate collisions in the 2D setting too. Here, we
aim to briefly consider this issue, while a detailed analysis will be a
subject of a separate work.

\begin{figure*}[t]
{\includegraphics[width=1.5\columnwidth]{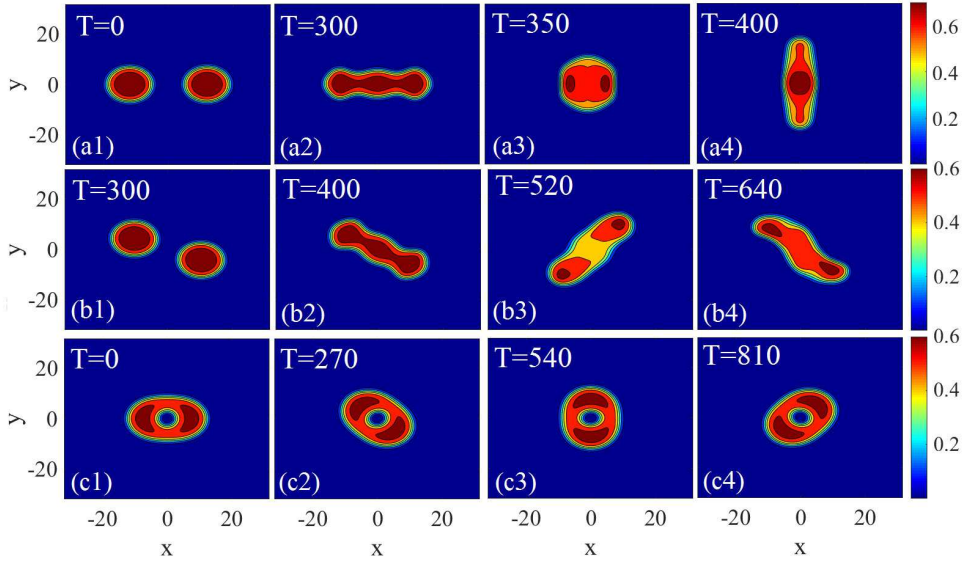}}
\caption{(a1)-(a4) Merger of two zero-vorticity QDs into a single droplet
with strongly oscillating eccentricity, initiated by $\protect\psi (\mathbf{r%
},t=0)=\protect\phi (x-x_{0},y)+\protect\phi (x+x_{0},y)$, with $x_{0}=12$
and norm $N=100$ of each QD. (b1)-(b4) Merger of two transversely kicked
zero-vorticity QDs, with $S=0$, into a rotating elongated droplet, initiated
by $\protect\psi (\mathbf{r},t=0)=\protect\phi (x-x_{0},y)e^{-iky}+\protect%
\phi (x+x_{0},y)e^{iky}$, with $x_{0}=12$, $N=100$, and kick $k=0.015$.
(c1)-(c4) Steady rotation of an oval-shaped vortex QD with vorticity $S=1$
and $N=200$. These results are produced by simulations of Eq. (\protect\ref%
{GPE}) with $L_{3}=0$.}
\label{dynamics}
\end{figure*}

Simulations, performed in the framework of Eq. (\ref{GPE}) with ${L_{3}=0}$,
demonstrate that the interaction between initially separated fundamental ($%
S=0$) QDs with zero phase difference leads, in the lossless model, to their
merger into a single prolate droplet (which is a natural outcome for the
fluid phase), that performs strong \textit{oscillations of eccentricity}, as
shown in Fig. \ref{dynamics}(a1)-(a4). This excited mode resembles
eccentricity oscillations performed by elliptically deformed circular kinks
in the 2D sine-Gordon model \cite{Denmark}.

Another option is the merger of two fundamental QDs, with opposite kicks
initially applied to them in the transverse ($y$) direction (in other words,
with \textit{torque }applied to the QD pair), as shown in Fig. \ref{dynamics}%
(b1)-(b4). This pair merges into a \textit{rotating} prolate droplet. A
different dynamical regime revealed by the simulations is rotation of an
oval-shaped vortex with embedded vorticity $S=\pm 1$, see Fig. \ref{dynamics}%
(c1)-(c4). The sign of embedded $S=+1$ or $-1$ correlates with the
counter-clockwise or clockwise direction of the rotation. The consideration
of these rotational regimes is suggested by recent studies of spinning
helium nanodroplets \cite{Science,PRB}.

Collisions between moving QDs were investigated too. The conclusions are
similar to what is known about other nonintegrable models \cite{Spain,BenLi}%
, including the above-mentioned model for one-dimensional QDs \cite{Grisha}:
fast moving QDs pass through each other quasi-elastically, while slowly
moving ones collide inelastically (not shown here). In particular, slowly
colliding QDs with $S=0$ merge into a single droplet, while slowly colliding
vortices suffer destruction.

\section{Conclusion}

The objective of this work is to investigate possibilities for the creation
of effectively two-dimensional self-trapped QDs (quantum droplets) in the
model based on nonlinearly coupled GPEs (Gross-Pitaevskii equations) for the
binary BEC, with self-repulsion in each component and dominating attraction
between them, the stabilization against the collapse being provided by the
LHY (Lee-Huang-Yang) effect. As recently demonstrated, the nonlinearity in
such an effective two-dimensional system takes the form of cubic terms
multiplied by the additional logarithmic factor. We have constructed
families of two-component QDs, with equal vorticities $S$, or opposite ones $%
\pm S$, embedded in each component, the latter species being called the HV
(hidden-vorticity) mode. While the entire family with $S=0$ is stable, an
essential finding is the stability region for the modes with the explicit
vorticity (identical in both components), $1\leq S\leq 5$ (while the recent
consideration of the vortex QDs in the 3D model reveals their stability
solely for $S=1$ and $2$ \cite{Leticia}). All the modes with $S\geq 6$ are
unstable in the region of norms, $N$, which are accessible for the analysis.
Both the existence and stability boundaries for the vortex QDs have been
identified, respectively, as $N_{\max }(S)$ and $N_{\mathrm{th}}(S)$. The
steep growth of the latter value, $N_{\mathrm{th}}(S)\sim S^{4}$, and a
still steeper scaling, $N_{\mathrm{th}}(S)\sim S^{6}$, predicted for
three-dimensional QDs, has been explained analytically, considering the
energy balance between the vortex mode and a set of fragments which may be
produced by its splitting. At the stability boundary, the modes with $S=1,2,3
$ seem as usual solitons with embedded vorticities, but deeper into the
stability area, all vortex QDs develop a flat-top shape. While this feature
may look similar to that known in models with competing self-focusing and
defocusing nonlinear terms (e.g., the cubic-quintic NLS equation in two
dimensions), the present system is essentially different, as it contains the
single nonlinear term, $\sim |\psi |^{2}\psi \ln \left( |\psi |^{2}\right) $%
, as mentioned above.

It was demonstrated here too that pivots of all modes with multiple
vorticities, $S\geq 2$, are subject to the structural instability, in the
sense that specially selected small perturbations may split the single pivot
into sets of $S$ or $S+2$ ones representing unitary vortices. However, the
structural instability remains virtually invisible, as it occurs in the
broad central \textquotedblleft hole" of the multiple-$S$ vortex, where
values of the fields are extremely small, and it does not cause any
dynamical instability of the states with $S\geq 2$.

The nearly-constant value of the density in the flat-top area, and radius $%
r_{\min }$ of the central \textquotedblleft hole" of the vortex mode (which
is asymptotically independent of $S$) have been identified in the
approximate analytical form, $r_{\min }$ being found by means of the
Thomas-Fermi approximation. The role of the three-body loss, which may be an
essential factor in the real experiment, was explored too, showing that it
does not preclude the possibility of the creation and observation of stable
vortex droplets (ones with relatively low initial norms may suddenly split
under the action of the loss). The existence and stability boundaries for
the QDs with all values of the explicit vorticity do not depend on the value
of the coupling constant, $g$, of the GPE system. A possible stability
region is also identified for HV modes, with vorticities $+1$ and $-1$ in
its two components. Unlike the QDs with the explicit vorticity, stability of
the HV states strongly depends on $g$, all states with $S_{+}=-S_{-}\geq 2$
being unstable. More complex robust dynamical modes were constructed and
briefly considered too, \textit{viz}., elliptically deformed QDs, with $S=0$
and $S=1$, which exhibit, severally, either strong eccentricity oscillations
or steady rotation.

These results may find realizations in the ongoing experiments with QDs,
some of which actually address nearly-2D configurations \cite%
{Cabrera2017,Cabrera2}. This possibility is quite important as, thus far, no
stable bright vortex solitons have been created experimentally in free space
or in any uniform physical medium. The vorticity may be imparted to the QD
by a laser beam carrying an orbital optical momentum \cite{impart}. The
necessary condition for that is that the droplet's size must be essentially
larger than the wavelength of light, which definitely holds in the present
setting.

In addition to the vortex states studied in this work, it may be relevant to
address excited 2D states with a more complex radial structure. It may be
also interesting to extend the systematic analysis to asymmetric
two-component systems, with unequal scattering lengths accounting for the
self-repulsion in the two components, and/or with unequal atomic masses
(heteronuclear mixtures \cite{preprint}), as well as with unequal numbers of
atoms in the two components. In particular, at finite temperature, the
imbalance may imitate a controllable thermal bath \cite{Cabrera2}.

\begin{acknowledgments}
We appreciate valuable discussions with F. Ancilotto and G. E.
Astrakharchik. This work was supported by NNSFC (China) through Grant No.
11874112, 11575063, by the joint program in physics between NSF and
Binational (US-Israel) Science Foundation through project No. 2015616, and
by the Israel Science Foundation through Grant No. 1286/17. B.A.M.
appreciated hospitality of the College of Electronic Engineering at the
South China Agricultural University (Guangzhou).
\end{acknowledgments}


\begin{thebibliography}{99}
\bibitem{GP1} L. P. Pitaevskii and S. Stringari, \textit{Bose-Einstein
Condensation} (Oxford University Press: Oxford, 2003).

\bibitem{GP2} C. J. Pethick and H. Smith, \textit{Bose--Einstein
Condensation in Dilute Gases} (Cambridge University Press: Cambridge, 2008).

\bibitem{review} B. A. Malomed, D. Mihalache, F. Wise, and L. Torner,
Spatiotemporal optical solitons, J. Optics B: Quant. Semicl. Opt. \textbf{7}%
, R53-R72 (2005);Viewpoint: On multidimensional solitons and their legacy in
contemporary Atomic, Molecular and Optical physics, J. Phys. B: At. Mol.
Opt. Phys. \textbf{49}, 170502 (2016).

\bibitem{Berge'} L. Berg\'{e}, Wave collapse in physics: principles and
applications to light and plasma waves, Phys. Rep. \textbf{303}, 259-370
(1998).

\bibitem{Fibich} G. Fibich, \textit{The Nonlinear Schr\"{o}dinger Equation:
Singular Solutions and Optical Collapse} (Springer: Heidelberg, 2015).

\bibitem{review2} B. A. Malomed, Multidimensional solitons: Well-established
results and novel findings, Eur. Phys. J. Special Topics \textbf{225},
2507-2532 (2016).

\bibitem{Spain} M. Quiroga-Teixeiro and H. Michinel, Stable azimuthal
stationary state in quintic nonlinear media, J. Opt. Soc. Am. B \textbf{14},
2004-2009 (1997).

\bibitem{Pego} R. L. Pego and H. A. Warchall, Spectrally stable encapsulated
vortices for nonlinear Schr\"{o}dinger equations, J. Nonlinear Sci. \textbf{%
12}, 347-394 (2002).

\bibitem{Kiev} T. A. Davydova and A. I. Yakimenko, Stable multicharged
localized optical vortices in cubic-quintic nonlinear media, J. Opt. A: Pure
Appl. Opt. \textbf{6}, S197-S201 (2004).

\bibitem{3D-vortex} D. Mihalache, D. Mazilu, L.-C. Crasovan, I. Towers, A.
V. Buryak, B. A. Malomed, L. Torner, J. P. Torres, and F. Lederer, Stable
spinning optical solitons in three dimensions, Phys. Rev. Lett. \textbf{88},
073902 (2002).

\bibitem{Cid} E. L. Falc\~{a}o-Filho, and C. B. de Ara\'{u}jo, G. Boudebs,
H. Leblond, and V. Skarka, Robust two-dimensional spatial solitons in liquid
carbon disulfide, Phys. Rev. Lett. \textbf{110}, 013901 (2013).

\bibitem{Cid2} A. S. Reyna, G. Boudebs, B. A. Malomed, and C. B. de Ara\'{u}%
jo, Robust self-trapping of vortex beams in a saturable optical medium,
Phys. Rev. A \textbf{93}, 013840 (2016).

\bibitem{Jqing2016} J. Qin, G. Dong, and B. A. Malomed, Stable giant vortex
annuli in microwave-coupled atomic condensates, Phys. Rev. A \textbf{94},
053611 (2016).

%
%

\bibitem{BenLi} H. Sakaguchi, B. Li, and B. A. Malomed, Creation of
two-dimensional composite solitons in spin-orbit-coupled self-attractive
Bose-Einstein condensates in free space, Phys. Rev. E \textbf{89}, 032920
(2014).

\bibitem{Luca} L. Salasnich, W. B. Cardoso, and B. A. Malomed, Localized
modes in quasi-two-dimensional Bose-Einstein condensates with spin-orbit and
Rabi couplings, Phys. Rev. A \textbf{90}, 033629 (2014).

\bibitem{Yongyao2017} Y. Li, Z. Luo, Y. Liu, Z. Chen, C. Huang, S. Fu, H.
Tan, and B. A. Malomed, Two-dimensional solitons and quantum droplets
supported by competing self- and cross-interactions in spin-orbit-coupled
condensates, New. J. Phys. \textbf{19}, 113043 (2017).

\bibitem{HanPu} Y.-C. Zhang, Z.-W. Zhou, B. A. Malomed, and H. Pu, Stable
solitons in three dimensional free space without the ground state:
Self-trapped Bose-Einstein condensates with spin-orbit coupling, Phys. Rev.
Lett. \textbf{115}, 253902 (2015).

\bibitem{SKA} S. Gautam and S. K. Adhikari, Three-dimensional vortex-bright
solitons in a spin-orbit-coupled spin-1 condensate, Phys. Rev. A \textbf{97}%
, 013629 (2018).

\bibitem{Jena} F. Eilenberger, K. Prater, S. Minardi, R. Geiss, U. R\"{o}%
pke, J. Kobelke, K. Schuster, H. Bartelt, S. Nolte, A. T\"{u}nnermann, and
T. Pertsch, Observation of discrete, vortex light bullets, Phys. Rev. X
\textbf{3}, 041031 (2013).

\bibitem{Petrov2015} D. S. Petrov, Quantum Mechanical Stabilization of a
collapsing Bose-Bose mixture, Phys. Rev. Lett. \textbf{115}, 155302 (2015).

\bibitem{Petrov2016} D. S. Petrov and G. E. Astrakharchik, Ultradilute
Low-Dimensional Liquids, Phys. Rev. Lett \textbf{117}, 100401 (2016).

\bibitem{LHY1957} T. D. Lee, K. Huang, and C. N. Yang, Eigenvalues and
eigenfunctions of a Bose system of hard spheres and its low-temperature
properties, Phys. Rev. \textbf{106}, 1135 (1957).

\bibitem{Sekino2018} Y. Sekino and Y. Nishida, Quantum droplet of
one-dimensional bosons with a three-body attraction, Phys. Rev. A \textbf{97}%
, 011602(R) (2018).

\bibitem{Randy} K. E. Strecker, G. B. Partridge, A. G. Truscott, and R. G.
Hulet, Formation and propagation of matter-wave soliton trains, Nature
\textbf{417}, 150-153 (2002).

\bibitem{Lev} L. Khaykovich, F. Schreck, G. Ferrari, T. Bourdel, J.
Cubizolles, L. D. Carr, Y. Castin, and C. Salomon, Formation of a
matter-wave bright soliton, Science \textbf{296}, 1290-1293 (2002).

\bibitem{Randy-review} K. E. Strecker, G. B. Partridge, A. G. Truscott, and
R. G. Hulet, Bright matter wave solitons in Bose--Einstein condensates, New
J. Phys. \textbf{5}, 73 (2003).

\bibitem{Cornish} S. L. Cornish, S. T. Thompson, and C. E. Wieman, Formation
of bright matter-wave solitons during the collapse of attractive
Bose-Einstein condensates, Phys. Rev. Lett. \textbf{96}, 170401 (2006).

\bibitem{Kasevich} P. Medley, M. A. Minar, N. C. Cizek, D. Berryrieser, and
M. A. Kasevich, Evaporative production of bright atomic solitons, Phys. Rev.
Lett. \textbf{112}, 060401 (2014).

\bibitem{Canberra} P. J. Everitt, M. A. Sooriyabandara, G. D. McDonald, K.
S. Hardman, C. Quinlivan, P. Manju, P. Wigley, J. E. Debs, J. D. Close, C.
C. N. Kuhn, and N. P. Robins, Observation of breathers in an attractive Bose
gas, arXiv:1509.06844

\bibitem{Aspect} S. Lepoutre, L. Fouch\'{e}, A. Boiss\'{e}, G. Berthet, G.
Salomon, A. Aspect, and T. Bourdel, Production of strongly bound $^{39}$K
bright solitons, Phys. Rev. A \textbf{94}, 053626 (2016).

\bibitem{Luca2} A. Cappellaro, T. Macr\'{\i}, G. F. Bertacco, and L.
Salasnich, Equation of state and self-bound droplet in Rabi-coupled Bose
mixtures, Sci. Rep. \textbf{7}, 13358 (2017).

\bibitem{Cabrera2017} C. R. Cabrera, L. Tanzi, J. Sanz, B. Naylor, P.
Thomas, P. Cheiney, L. Tarruell, Quantum liquid droplets in a mixture of
Bose-Einstein condensates, Science \textbf{359}, 301-304 (2018).

\bibitem{comment} I. Ferrier-Barbut and T. Pfau, Quantum liquids get thin,
Science \textbf{359}, 274-275 (2018).

\bibitem{Cabrera2} P. Cheiney, C. R. Cabrera, J. Sanz, B. Naylor, L. Tanzi,
and L. Tarruell, Bright soliton to quantum droplet transition in a mixture
of Bose-Einstein condensates, Phys. Rev. Lett. \textbf{120}, 135301 (2018).

\bibitem{Inguscio} G. Semeghini, G. Ferioli, L. Masi, C. Mazzinghi, L.
Wolswijk, F. Minardi, M. Modugno, G. Modugno, M. Inguscio, and M. Fattori,
Self-bound quantum droplets in atomic mixtures, Phys. Rev. Lett. \textbf{120}%
, 235301 (2018).

\bibitem{FR1} C. D'Errico, M. Zaccanti, M. Fattori, G. Roati, M. Inguscio,
G. Modugno, and A. Simoni, Feshbach resonances in ultracold $^{39}$K, New J.
Phys. \textbf{9}, 223 (2007).

\bibitem{FR2} M. Lysebo and L. Veseth, Feshbach resonances and transition
rates for cold homonuclear collisions between $^{39}$K and $^{41}$K atoms,
Phys. Rev. A \textbf{81}, 032702 (2010).

\bibitem{Barbut2016} I. Ferrier-Barbut, H. Kadau, M. Schmitt, M. Wenzel, and
T. Pfau, Observation of quantum droplets in a strongly dipolar Bose gas,
Phys. Rev. Lett. \textbf{116}, 215301 (2016).

\bibitem{Schmitt2016} M. Schmitt, M. Wenzel, F. B\"{o}ttcher, I.
Ferrier-Barbut, and T. Pfau, Self-bound droplets of a dilute magnetic
quantum liquid, Nature \textbf{539}, 259-262 (2016).

\bibitem{PRX} L. Chomaz, S. Baier, D. Petter, M. J. Mark, F. W\"{a}chtler,
L. Santos, and F. Ferlaino, Quantum-fluctuation-driven crossover from a
dilute Bose-Einstein condensate to a macrodroplet in a dipolar quantum
fluid, Phys. Rev. X \textbf{6}, 041039 (2016).

\bibitem{Lima2011} A. R. P. Lima, and A. Pelster, Quantum fluctuations in
dipolar Bose gases, Phys. Rev. A \textbf{84}, 041604(R) (2011).

\bibitem{Saito2016} H. Saito, Path-integral Monte Carlo study on a droplet
of a dipolar Bose-Einstein condensate stabilized by quantum fluctuation, J.
Phys. Soc. Jpn. \textbf{85}, 053001 (2016).

\bibitem{Xi} K.-T. Xi and H. Saito, Droplet formation in a Bose-Einstein
condensate with strong dipole-dipole interaction, Phys. Rev. A \textbf{93},
011604(R) (2016).

\bibitem{Bisset2016} R. N. Bisset, R. M. Wilson, D. Baillie, and P. B.
Blakie, Ground-state phase diagram of a dipolar condensate with quantum
fluctuations, Phys. Rev. A \textbf{94}, 033619 (2016).

\bibitem{Wachtler2016} F. W\"{a}chtler and L. Santos, Quantum filaments in
dipolar Bose-Einstein condensate, Phys. Rev. A \textbf{93}, 061603(R) (2016).

\bibitem{Pastukov2017} V. Pastukhov, Beyond mean-field properties of binary
dipolar Bose mixtures at low temperatures, Phys. Rev. A \textbf{95}, 023614
(2017)

\bibitem{Cinti2017} F. Cinti and M. Boninsegni, Classical and quantum
filaments in the ground state of trapped dipolar Bose gases, Phys. Rev. A
\textbf{96}, 013627 (2017).

\bibitem{Edler2017} D. Edler, C. Mishra, F. W\"{a}chtler, R. Nath, S. Sinha,
and L. Santos, Quantum fluctuations in quasi-one-dimensional dipolar
Bose-Einstein condensates, Phys. Rev. Lett. \textbf{119}, 050403 (2017).

\bibitem{Sadhan} S. K. Adhikari, Statics and dynamics of a self-bound
dipolar matter-wave droplet, Laser Phys. Lett. \textbf{14}, 025501 (2017).

\bibitem{Macri} A. Cidrim, F. E. A. dos Santos, E. A. L. Henn, and T. Macr%
\'{\i}, Vortices in self-bound dipolar droplets, Phys. Rev. A \textbf{98},
023618 (2018).

\bibitem{Leticia} Y. V. Kartashov, B. A. Malomed, L. Tarruell, and L.
Torner, Three-dimensional droplets of swirling superfluids, Phys. Rev. A
\textbf{98}, 013612 (2018).

\bibitem{Zhengwei2013} W. Zheng, Z. Yu, X. Cui, and H. Zhai, Properties of
Bose gases with the Raman-induced spin-orbit coupling, J. Phys. B: At. Mol.
Opt. Phys. \textbf{46}, 134007 (2013).



\bibitem{hidden} M. Brtka, A. Gammal, and B. A. Malomed, Hidden vorticity in
binary Bose-Einstein condensates, Phys. Rev. A \textbf{82}, 053610 (2010).

\bibitem{IT1} D. L. Feder, M. S. Pindzola, L. A. Collins, B. I. Schneider,
and C. W. Clark, Dark-soliton states of Bose-Einstein condensates in
anisotropic traps, Phys. Rev. A \textbf{62}, 053606 (2000).

\bibitem{IT2} W. Z. Bao and Q. Du, Computing the ground state solution of
Bose-Einstein condensates by a normalized gradient flow, SIAM\ J. Scientific
Computing \textbf{25}, 1674-1697 (2004).

\bibitem{VS} D. Mihalache, D. Mazilu, B. A. Malomed, and F. Lederer, Vortex
stability in nearly-two-dimensional Bose-Einstein condensates with
attraction, Phys. Rev. A \textbf{73}, 043615 (2006).

\bibitem{VK} M. Vakhitov and A. Kolokolov, Stationary solutions of the wave
equation in a medium with nonlinearity saturation, Radiophys. Quantum
Electron. \textbf{16}, 783-789 (1973).

\bibitem{Pushka} Kh. I. Pushkarov, D. I. Pushkarov, and I. V. Tomov,
Self-action of light beams in nonlinear media: soliton solutions, Opt.
Quantum Electron. \textbf{11}, 471-478 (1979).

\bibitem{Enns} S. Cowan, R. H. Enns, S. S. Rangnekar, and S. S. Sanghera,
Quasi-soliton and other behavior of the nonlinear cubic-quintic Schr\"{o}%
dinger equation, Can. J. Phys. \textbf{64}, 311-315 (1986).

\bibitem{Grisha} G. E. Astrakharchik and B. A. Malomed, Dynamics of
one-dimensional quantum droplets, Phys. Rev. A \textbf{98}, 013631 (2018).

\bibitem{Luca3} A. Cappellaro, T. Macr\`{\i}, and L. Salasnich1, Collective
modes across the soliton-droplet crossover in binary Bose mixtures, Phys.
Rev. A \textbf{97}, 053623 (2018).

\bibitem{new} X. Cui, Spin-orbit-coupling-induced quantum droplet in
ultracold Bose-Fermi mixtures, Phys. Rev. A \textbf{98}, 023630 (2018).

\bibitem{Boronat} V. Cikojevi\'{c}, K. D\v{z}elalija, P. Stipanovi\'{c}, and
L. Vranje\v{s} Marki\'{c}, Ultradilute quantum liquid drops, Phys. Rev. B
\textbf{97}, 140502(R) (2018).

\bibitem{Zillich} C. Staudinger, F. Mazzanti, and R. E. Zillich, Self-bound
Bose mixtures, Phys. Rev. A \textbf{98}, 023633 (2018).

\bibitem{Chiquillo} E. Chiquillo, Equation of state of the one- and
three-dimensional Bose-Bose gases, Phys. Rev. A 97, 063605 (2018).

\bibitem{Feder} A. L. Fetter, Rotating trapped Bose-Einstein condensates,
Rev. Mod. Phys. \textbf{81}, 647-691 (2009).

\bibitem{Shlyap} Y. Kagan, B. V. Svistunov, and G. V. Shlyapnikov, Effect of
Bose condensation on inelastic processes in gases, JETP Lett. \textbf{42},
209-212 (1985).

\bibitem{T} A. Boudjem\^{a}a, Quantum dilute droplets of dipolar bosons at
finite temperature, Ann. Phys. (N.Y.) \textbf{381}, 69-79 (2017).

\bibitem{Nick} \textit{Quantum Gases: Finite Temperature and Non-Equilibrium
Dynamics}, ed. by N. Proukakis, S. Gardiner, M. Davis, and M. Szyma\'{n}ska
(Imperial College Press: London, 2013).

\bibitem{K-39} G. Modugno, G. Ferrari, G. Roati, R. J. Brecha, A. Simoni,
and M. Inguscio, Bose-Einstein condensation of potassium atoms by
sympathetic cooling, Science \textbf{294}, 1320-1322 (2001).

\bibitem{Denmark} P. L. Christiansen, N. Gr\o nbech-Jensen, P. S. Lomdahl,
and B. A. Malomed, Oscillations of eccentric pulsons, Physica Scripta
\textbf{55}, 131-134 (1997).

\bibitem{Science} L. F. Gomez, K. R. Ferguson, J. P. Cryan, C. Bacellar, R.
M. P. Tanyag, C. Jones, S. Schorb, D. Anielski, A. Belkacem, C. Bernando, R.
Boll, J. Bozek, S. Carron, G. Chen, T. Delmas, L. Englert, S. W. Epp, B.
Erk, L. Foucar, R. Hartmann, A. Hexemer, M. Huth, J. Kwok, S. R. Leone, J.
H. S. Ma, F. R. N. C. Maia, E. Malmerberg, S. Marchesini, D. M. Neumark, B.
Poon, J. Prell, D. Rolles, B. Rudek, A. Rudenko, M. Seifrid, K. R.
Siefermann, F. P. Sturm, M. Swiggers, J. Ullrich, F. Weise, P. Zwart, C.
Bostedt, O. Gessner, and A. F. Vilesov, Shapes and vorticities of superfluid
helium nanodroplets, Science \textbf{345}, 906-909 (2014).

\bibitem{PRB} F. Ancilotto, M. Barranco, and M. Pi, Spinning superfluid 4 He
nanodroplets, Phys. Rev. B \textbf{97}, 184515 (2018).

\bibitem{impart} M. F. Andersen, C. Ryu, P. Clad\'{e}, V. Natarajan, A.
Vaziri, K. Helmerson, and W. D. Phillips, Quantized rotation of atoms from
photons with orbital angular momentum, Phys. Rev. Lett. \textbf{97}, 170406
(2006); R. Pugatch, M. Shuker, O. Firstenberg, A. Ron, and N. Davidson,
Topological stability of stored optical vortices \textit{ibid}. \textbf{98},
203601 (2007).

\bibitem{preprint} F. Ancilotto, M. Barranco, M. Guilleumas, and M. Pi,
Self-bound ultra dilute Bose mixtures within local density approximation,
arXiv:1808.09197.
\end{thebibliography}
\end{document}